\documentclass[twocolumn,superscriptaddress,showkeys]{revtex4}
\usepackage{amssymb,epsf,epsfig,amsmath,color}
\usepackage{multirow}
\newcommand{\lsim}{
\mathrel{\hbox{\rlap{\hbox{\lower4pt\hbox{$\sim$}}}\hbox{$<$}}}}
\newcommand{\gsim}{
\mathrel{\hbox{\rlap{\hbox{\lower4pt\hbox{$\sim$}}}\hbox{$>$}}}}

\def\D0{D\O }
\def\theabstract{We have just entered an era of precision measurements for $B_s$-decay 
observables. A characteristic feature of the $B_s$-meson system is $B^0_s$--$\bar B^0_s$ 
mixing, which exhibits a sizable decay width difference. The latter feature leads to a 
subtle complication for the extraction of branching ratios of $B_s$ decays from 
untagged data samples, leading to systematic biases as large as ${\cal O}(10\%)$ 
that depend on the dynamics of the considered decay. We point out that this effect 
can only be corrected for using information from a time-dependent analysis and suggest
the use of the effective $B_s$ decay lifetime, which can already be extracted from
the untagged data sample, for this purpose. 
We also address several experimental issues that can play a role in the extraction
of effective lifetimes at a hadron collider, and advocate the use of the $B_s$ branching 
ratios, as presented in this note, for consistent comparisons of theoretical
calculations and experimental measurements in particle listings.}

\begin{document}
\begin{titlepage}
\vspace*{1.7truecm}
\begin{flushright}
Nikhef-2012-005
\end{flushright}

\vspace{1.6truecm}

\begin{center}
\boldmath
{\Large{\bf Branching Ratio Measurements of $B_s$ Decays}}
\unboldmath
\end{center}

\vspace{1.2truecm}

\begin{center}
{\bf Kristof De Bruyn\,${}^a$, Robert Fleischer\,${}^{a,b}$, Robert Knegjens\,${}^a$, \\
Patrick Koppenburg\,${}^a$, Marcel Merk\,${}^{a,b}$, Niels Tuning\,${}^a$}

\vspace{0.5truecm}

${}^a${\sl Nikhef, Science Park 105, NL-1098 XG Amsterdam, The Netherlands}

${}^b${\sl  Department of Physics and Astronomy, Vrije Universiteit Amsterdam,\\
NL-1081 HV Amsterdam, The Netherlands}

\end{center}

\vspace*{2.7cm}

\begin{center}
\large{\bf Abstract}\\

\vspace*{0.6truecm}

\begin{tabular}{p{13.5truecm}}
\theabstract
\end{tabular}

\end{center}

\vspace*{2.7truecm}

\vfill

\noindent
April 2012

\end{titlepage}

\newpage
\thispagestyle{empty}
\mbox{}

\newpage
\thispagestyle{empty}
\mbox{}

\rule{0cm}{23cm}

\newpage
\thispagestyle{empty}
\mbox{}

\setcounter{page}{0}

\preprint{Nikhef-2012-nnn}

\date{\today}

\title{\boldmath Branching Ratio Measurements of $B_s$ Decays \unboldmath}

\author{Kristof De Bruyn}
\affiliation{Nikhef, Science Park 105, NL-1098 XG Amsterdam, The Netherlands}

\author{Robert Fleischer}
\affiliation{Nikhef, Science Park 105, NL-1098 XG Amsterdam, The Netherlands}
\affiliation{Department of Physics and Astronomy, Vrije Universiteit Amsterdam,
NL-1081 HV Amsterdam, The Netherlands}

\author{Robert Knegjens}
\affiliation{Nikhef, Science Park 105, NL-1098 XG Amsterdam, The Netherlands}

\author{Patrick Koppenburg}
\affiliation{Nikhef, Science Park 105, NL-1098 XG Amsterdam, The Netherlands}

\author{Marcel Merk}
\affiliation{Nikhef, Science Park 105, NL-1098 XG Amsterdam, The Netherlands}
\affiliation{Department of Physics and Astronomy, Vrije Universiteit Amsterdam,
NL-1081 HV Amsterdam, The Netherlands}

\author{Niels Tuning}
\affiliation{Nikhef, Science Park 105, NL-1098 XG Amsterdam, The Netherlands}

\begin{abstract}
\vspace{0.2cm}\noindent
\theabstract
\end{abstract}


\maketitle

\section{Introduction}
Weak decays of $B_s$ mesons encode valuable information for the exploration of the
Standard Model (SM). The simplest observables are branching ratios, which give the probability 
of the considered decay to occur. Measurements of $B_s$ branching ratios at hadron colliders, 
such as Fermilab's Tevatron and CERN's Large Hadron Collider (LHC), would require 
knowledge of the $B_s$ production cross-section, which presently makes absolute 
branching ratio measurements impossible. Hence experimental control channels and 
the ratio of the $f_s/f_{u,d}$ fragmentation functions, describing the probability that a 
$b$ quark hadronizes as a $\bar B_q$ meson \cite{FST}, are required for the conversion 
of the observed number of decays into the branching ratio. At $e^+e^-$ $B$ factories operated
at the $\Upsilon(5S)$ resonance, the total number of produced $B_s$ mesons is measured
separately and subsequently 
also allows for the extraction of the $B_s$ branching ratio from the data~\cite{Drutskoy:2006xc}.

A key feature of the $B_s$ mesons is $B^0_s$--$\bar B^0_s$ mixing, which leads
to quantum-mechanical, time-dependent oscillations between the $B^0_s$ and 
$\bar B^0_s$ states. 
In contrast to the $B_d$ system, the $B_s$ mesons exhibit a sizable difference between 
the decay widths of the light and heavy mass eigenstates, $\Gamma_{\rm L}^{(s)}$ and 
$\Gamma_{\rm H}^{(s)}$, respectively \cite{LN}. Currently the most precise measurement 
is extracted from the $B^0_s \to J/\psi \phi$ channel by the LHCb 
collaboration \cite{LHCb-Mor-12}:
\begin{equation}\label{ys-LHCb}
	y_s \equiv \frac{\Delta\Gamma_s}{2\,\Gamma_s}\equiv
	\frac{\Gamma_{\rm L}^{(s)} - \Gamma_{\rm H}^{(s)}}{2\,\Gamma_s}= 0.088 \pm 0.014;
\end{equation}
$\tau_{B_s}^{-1} \equiv \Gamma_s \equiv \bigl[\Gamma_{\rm L}^{(s)} + 
\Gamma_{\rm H}^{(s)}\bigr]/2=\left(0.6580 \pm 0.0085\right)\mbox{ps}^{-1}$
is the inverse of the $B_s$ mean lifetime $\tau_{B_s}$.

In view of the sizable decay width difference, Eq.~\eqref{ys-LHCb}, special care has to
be taken when dealing with the concept of a branching ratio. We shall clarify this 
issue and give an expression, allowing us to convert the experimentally measured 
$B_s$ branching ratio into the corresponding ``theoretical" branching ratio.
The latter is not affected by $B^0_s$--$\bar B^0_s$ mixing and encodes the information for the 
comparison with branching ratios of $B^0_d$ decays, where the relative decay 
width difference at the $10^{-3}$ level \cite{LN} can be neglected, or branching ratios 
of $B_u^+$ modes. 

The difference between these two branching ratio concepts involves $y_s$ and is specific 
for the considered $B_s$ decay, thereby involving non-perturbative parameters.
However, measuring the
effective lifetime of the considered $B_s$ decay, the effect can be included in a 
clean way.

In experimental analyses, this subtle effect has so far been neglected or only been 
partially addressed; examples are the branching ratio measurements of the 
$B_s\to K^+ K^-$~\cite{BRKK}, 
$B_s\to J/\psi f_0(980)$~\cite{Jpsif0},  $B_s\to J/\psi K_{\rm S}$~\cite{PsiKs},
$B_s\to D_s^{+}D_s^{-}$~\cite{DsDs} and
$B^0_s\to D_s^{-}\pi^{+}$~\cite{Dspi} decays by the LHCb, CDF, \D0 and Belle
collaborations.

\section{Experiment Versus Theory}
What complicates the concept of a $B_s$ branching ratio
is the fact that the untagged decay rate is the sum of two exponentials \cite{DFN}:
\begin{align}
	\langle \Gamma(B_s(t)\to f)\rangle
	&\equiv\ \Gamma(B^0_s(t)\to f)+ \Gamma(\bar B^0_s(t)\to f) \notag\\
	&=\  R^f_{\rm H} e^{-\Gamma_{\rm H}^{(s)} t} + R^f_{\rm L} e^{-\Gamma_{\rm L}^{(s)} t},
	\label{untagged}
\end{align}
corresponding to two mass eigenstates with different lifetimes.
Using Eq.~(\ref{ys-LHCb}), we write
\begin{align}
	&\langle \Gamma(B_s(t)\to f)\rangle
	= \left(R^f_{\rm H} + R^f_{\rm L}\right) e^{-\Gamma_s\,t} \notag\\
	&\times\left[ \cosh\left(\frac{y_s\, t}{\tau_{B_s}}\right)+
	{\cal A}^f_{\rm \Delta\Gamma}\,\sinh\left(\frac{y_s\, t}
	{\tau_{B_s}}\right)\right],
	\label{untagged2}
\end{align}
where 
\begin{equation}
{\cal A}^f_{\Delta\Gamma} \equiv \frac{R^f_{\rm H} - R^f_{\rm L}}{R^f_{\rm H} + R^f_{\rm L}}
\label{defADG}
\end{equation}
is a final-state dependent observable.

In experiment it is common practice to extract a branching ratio from 
the total event yield, ignoring information on the particles' lifetime. The ``experimental"
branching ratio can thus be defined as follows~\cite{DFN}:
\begin{align}
	&{\rm BR}\left(B_s \to f\right)_{\rm exp} 
	\quad\equiv \frac{1}{2}\int_0^\infty \langle \Gamma(B_s(t)\to f)\rangle\, dt 
	\label{defBrExp}\\
	&\quad= \frac{1}{2}\left[ \frac{R^f_{\rm H}}{\Gamma^{(s)}_{\rm H}} + 
	\frac{R^f_{\rm L}}{\Gamma^{(s)}_{\rm L}}\right]
	= \frac{\tau_{B_s}}{2}\left(R^f_{\rm H} + R^f_{\rm L}\right)
	\left[\frac{1 + {\cal A}^f_{\Delta\Gamma}\, y_s}{1-y_s^2} \right].\nonumber
	\end{align}
Note that this quantity is the average of the branching ratios for the heavy and light 
mass eigenstates. 

On the other hand, what is generally calculated theoretically are CP-averaged decay rates 
in the flavor-eigenstate basis, i.e.\
\begin{equation}
	\langle \Gamma(B_s(t)\to f)\rangle\big|_{t=0}
	= \Gamma(B^0_s\to f)+ \Gamma(\bar B^0_s\to f).
\end{equation}
This leads to the following definition of the ``theoretical" branching ratio:
\begin{align}
	{\rm BR}\left(B_s \to f\right)_{\rm theo} &\equiv 
	\frac{\tau_{B_s}}{2}\langle \Gamma(B^0_s(t)\to f)\rangle\Big|_{t=0}
	 \nonumber\\
	 &= \frac{\tau_{B_s}}{2}\left(R^f_{\rm H} + R^f_{\rm L}\right).
	 \label{defBrTheo}
\end{align}
By considering $t=0$, the effect of $B^0_s$--$\bar B^0_s$ mixing is ``switched off".
The advantage of this $B_s$ branching ratio definition, which has been used, for instance 
in Refs.~\cite{RF-BsJpsiKS,FFM}, is that it allows a straightforward comparison with 
branching ratios of $B^0_d$ or $B_u^+$ mesons by means of the $SU(3)$ flavor 
symmetry of strong interactions.

The experimentally measurable branching ratio, Eq.~(\ref{defBrExp}), can be converted into
the ``theoretical" branching ratio defined by Eq.~(\ref{defBrTheo}) through
\begin{equation}\label{BRratio}
	{\rm BR}\left(B_s \to f\right)_{\rm theo}
	= \left[\frac{1-y_s^2}{1 + {\cal A}^f_{\Delta\Gamma}\, y_s}\right]
	{\rm BR}\left(B_s \to f\right)_{\rm exp}.
\end{equation}
In the case of $y_s=0$, the theoretical and experimental branching ratio definitions are equal.

\begin{figure}[!t]
  \begin{center}
    \begin{picture}(250,170)(0,0)
      \put(0,0){\includegraphics[scale=0.33]{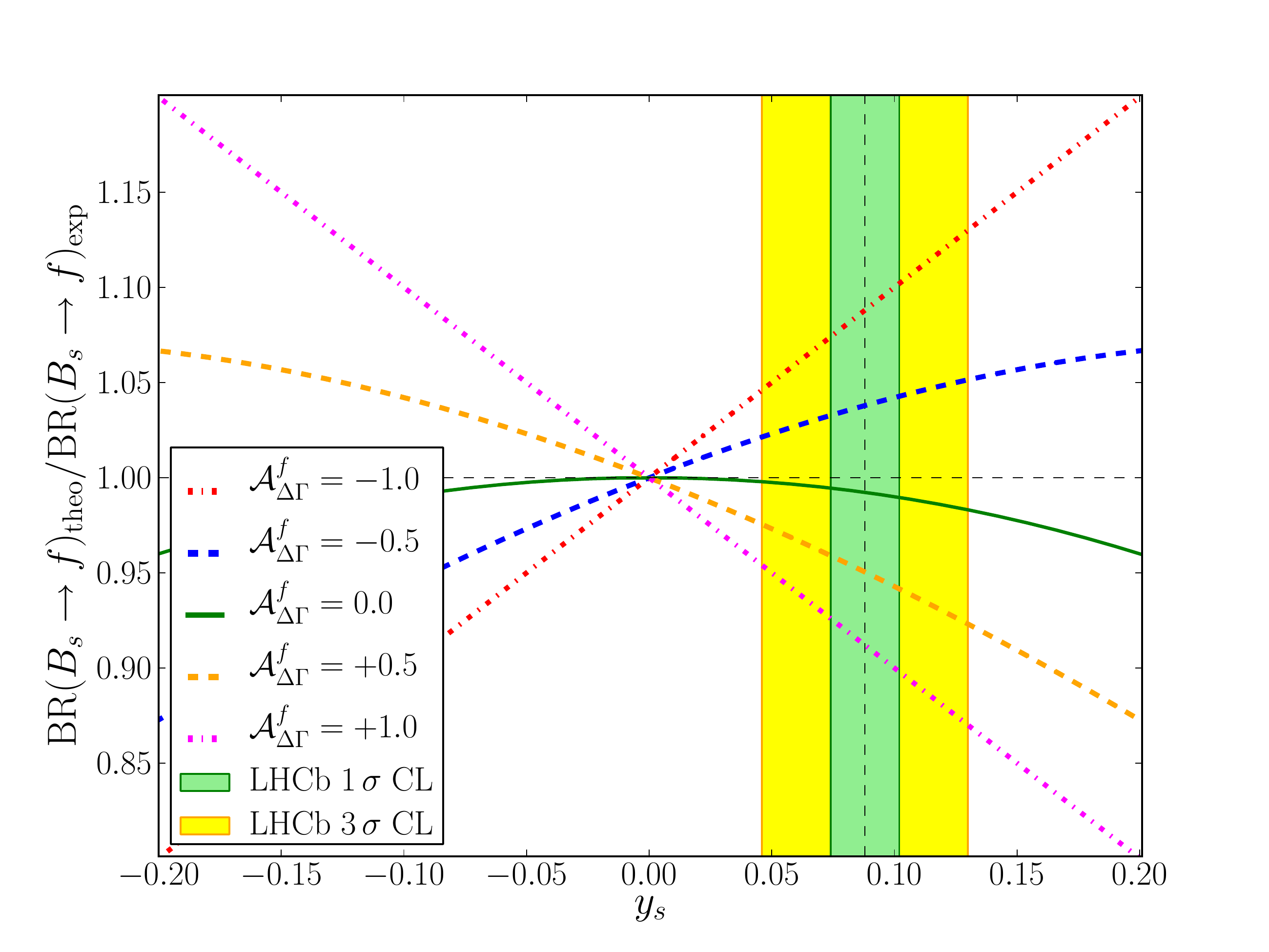}}
    \end{picture}
    \caption[dG/dq2]{\em Illustration of Eq.~(\ref{BRratio}) 
    for various values of ${\cal A}_{\Delta\Gamma}^f$.
    We also show the current LHCb measurement of $y_s$~\cite{LHCb-Mor-12}.}\label{fig:fig1}
  \end{center}
\end{figure}

Inspection of Eq.~\eqref{BRratio} reveals that $y_s$ and ${\cal A}_{\Delta\Gamma}^f$ 
are required for the translation of the experimental  branching ratios
into their theoretical counterparts. Ideally, the latter quantities should eventually
be used in particle compilations, in our opinion.

The decay width parameter $y_s$ is universal and has already been measured, as 
summarized in Eq.~(\ref{ys-LHCb}). In Fig.~\ref{fig:fig1}, we illustrate Eq.~\eqref{BRratio} for a 
variety of values of  ${\cal A}_{\Delta\Gamma}^f$ and observe that differences 
between ${\rm BR}\left(B_s \to f\right)_{\rm theo}$ and ${\rm BR}\left(B_s \to f\right)_{\rm exp}$ 
as large as ${\cal O}(10\%)$ may arise.

The simplest situation corresponds to flavor-specific (FS) decays such as $B^0_s\to D_s^-\pi^+$, 
where ${\cal A}^{\rm FS}_{\Delta\Gamma}=0$ and the correction factor is simply 
given by $1-y_s^2$. 

However, if both the $B^0_s$ and the $\bar B^0_s$ mesons can decay into the final state $f$, 
the observable ${\cal A}^f_{\Delta\Gamma}$ is more involved and depends, in general,
on non-perturbative hadronic parameters, CP-violating weak decay phases, and the 
$B^0_s$--$\bar B^0_s$ mixing phase $\phi_s$. Assuming the SM structure for the
decay amplitudes and using the $SU(3)$ flavor symmetry to determine the 
hadronic parameters from relations to $B_d$ decays, theoretical analyses of 
${\cal A}^f_{\Delta\Gamma}$ were performed for the final states 
$J/\psi\, \phi$ \cite{FFM}, $K^+ K^-$~\cite{Fleischer:2010ib}, 
$J/\psi\, f_0(980)$~\cite{Fleischer:2011au}, $J/\psi\, K_{\rm S}$~\cite{DeBruyn:2010hh} 
and $D_s^+D_s^-$~\cite{Fleischer:2007zn}.

\section{Using Lifetime Information}
\begin{center}
\begin{table*}[!t]
\begin{ruledtabular} 
     \begin{tabular}{lrrcc}
     \multirow{2}{*}{$B_s \to f$} & ${\rm BR}(B_s \to f)_{\rm exp}$\phantom{phm} & \multirow{2}{*}{${\cal A}_{\Delta\Gamma}^f(\text{SM})$\phantom{phantm}} & \multicolumn{2}{c}{${\rm BR}\left(B_s \to f\right)_{\rm theo}/{\rm BR}\left(B_s \to f\right)_{\rm exp}$} \\
      & (measured)\phantom{phm} & & From Eq.\ \eqref{BRratio} & From Eq.\ \eqref{BRratioT} \\
      \hline
      $ J/\psi f_0(980)$ & $(1.29^{+0.40}_{-0.28}) \times 10^{-4}$ \cite{PDG} & $0.9984\pm0.0021$ \cite{Fleischer:2011au} & $0.912\pm0.014$ & $0.890\pm0.082$ \cite{Jpsif0}\phantom{0}\\
      $ J/\psi K_{\rm S}$ & $(3.5 \pm 0.8) \times 10^{-5}$ \cite{PsiKs}\phantom{1} & $0.84\pm0.17$\phantom{00} \cite{DeBruyn:2010hh} & $0.924\pm0.018$ & N/A  \\
      $ D_s^-\pi^+$ & $(3.01 \pm 0.34) \times 10^{-3}$ \cite{Dspi}\phantom{1} & 0 (exact)\phantom{phantm} & $0.992\pm0.003$ & N/A  \\
      $ K^+ K^-$ & $(3.5 \pm 0.7) \times 10^{-5}$ \cite{PDG} & $-0.972\pm0.012$\phantom{0} \cite{Fleischer:2010ib} & $1.085\pm0.014$ & $1.042\pm0.033$ \cite{tauKK}\\
      $ D_s^+D_s^-$ & $(1.04^{+0.29}_{-0.26}) \times10^{-2}$ \cite{PDG} & $-0.995\pm0.013$\phantom{0} \cite{Fleischer:2007zn} & $1.088\pm0.014$ & N/A \\
     \end{tabular}
\end{ruledtabular}
    \caption{Factors for converting 
      ${\rm BR}\left(B_s \to f\right)_{\rm exp}$ (see \eqref{defBrExp}) into 
      ${\rm BR}\left(B_s \to f\right)_{\rm theo}$ (see \eqref{defBrTheo}) by means of 
      Eq.~\eqref{BRratio} with theoretical estimates for ${\cal A}^f_{\Delta\Gamma}$.
 Whenever effective lifetime information is available, the corrections are also calculated 
 using Eq.~\eqref{BRratioT}. }\label{tab:BRcomp}
\end{table*}
\end{center}

The simplest possibility for implementing Eq.~(\ref{BRratio}) is to use theoretical information
about the ${\cal A}^f_{\Delta\Gamma}$ observables. However, this input can be avoided
once time information of the untagged $B_s$ decay data sample becomes available. Then
the effective lifetime of the $B_s\to f$ decay can be determined, which is theoretically defined as the 
time expectation value of the untagged rate~\cite{Fleischer:2011cw}:
\begin{align}
	&\tau_f \equiv \frac{\int_0^\infty t\,\langle \Gamma(B_s(t)\to f)\rangle\, dt}
	{\int_0^\infty \langle \Gamma(B_s(t)\to f)\rangle\, dt} \notag\\ 
	&\quad = \frac{\tau_{B_s}}{1-y_s^2}\left[\frac{1+2\,{\cal A}^f_{\Delta\Gamma}y_s + y_s^2}
	{1 + {\cal A}^f_{\Delta\Gamma} y_s}\right].
	\label{effLifetime}
\end{align}
The advantage of $\tau_f$ is that it allows an efficient extraction of the product of 
${\cal A}^f_{\Delta\Gamma}$ and $y_s$. Using the effective lifetime, 
Eq.~\eqref{BRratio} can be expressed as
\begin{equation}
	{\rm BR}\left(B_s \to f\right)_{\rm theo}
	= \left[2 - \left(1-y_s^2\right)\frac{\tau_f}{\tau_{B_s}}\right]{\rm BR}\left(B_s \to f\right)_{\rm exp}.
	\label{BRratioT}
\end{equation}
Note that on the right-hand side of this equation only measurable quantities 
appear and that the decay width difference $y_s$ enters at second order. The measurement
of effective lifetimes is hence not only an interesting topic for obtaining constraints on the
$B^0_s$--$\bar B^0_s$ mixing parameters \cite{Fleischer:2011cw}, but an integral part of the
determination of the ``theoretical" $B_s$ branching ratios from the  data. 

In Table~\ref{tab:BRcomp}, we list the correction factors for converting the experimentally
measured branching ratios 
into the theoretical branching ratios for various decays. 
Here we have used theoretical 
information for ${\cal A}_{\Delta\Gamma}^f$ and Eq.~\eqref{BRratio}, or -- if available --
the effective decay lifetimes and Eq.~\eqref{BRratioT}. 

The rare decay $B^0_s\to \mu^+\mu^-$, which is very sensitive to New Physics 
\cite{buras}, is also affected by $\Delta\Gamma_s$.  In Ref.~\cite{Bsmumu}, 
we give a detailed discussion of this key $B_s$ decay, showing that the helicities of the
muons need not be measured to deal with this problem, and that $\Delta\Gamma_s$ actually
offers a new window for New Physics  in $B^0_s\to \mu^+\mu^-$.

\boldmath
\section{$B_s\to VV$ Decays}
\unboldmath
Another application is given by $B_s$ transitions into two vector mesons, 
such as $B_s\to J/\psi\phi$ \cite{JpsiPhi}, 
$B_s\to K^{*0}\bar K^{*0}$ \cite{LHCbKstarKstar} and $B_s\to D_s^{*+}D_s^{*-}$~\cite{DsDs}. 
Here an angular analysis of the decay products of the vector mesons has to be performed 
to disentangle the CP-even and CP-odd final states, which affects the branching 
fraction determination in a subtle way, as recognized in Refs.~\cite{LHCbKstarKstar,virto}.
Using linear polarization states 
$0, \parallel$ with CP eigenvalue $\eta_k=+1$ and $\perp$ with CP eigenvalue $\eta_k=-1$
\cite{rosner}, the generalization of Eq.~(\ref{BRratio}) is given by
\begin{equation}
\mbox{BR}_{\rm theo}^{VV}=\left(1-y_s^2\right) 
\left[ \sum_{k=0,\parallel,\perp}  \frac{f_{VV,k}^{\rm exp}}{1+y_s{\cal A}_{\Delta\Gamma}^{VV,k}}\right]
\mbox{BR}_{\rm exp}^{VV},
\end{equation}
where $f_{VV,k}^{\rm exp}={\rm BR}^{VV,k}_{\rm exp}/{\rm BR}^{VV}_{\rm exp}$
and $\mbox{BR}^{VV}_{\rm exp}\equiv \sum_k  \mbox{BR}_{\rm exp}^{VV,k}$
so that $\sum_k f_{VV,k}^{\rm exp} =1$. 
As discussed in Ref.~\cite{Fleischer:2011cw}, 
assuming the SM structure for the decay amplitudes, we can write
\begin{equation}
{\cal A}_{\Delta\Gamma}^{VV,k}=-\eta_k\sqrt{1-C_{VV,k}^2}\cos(\phi_s+\Delta\phi_{VV,k}),
\end{equation}
where $C_{VV,k}$ describes direct CP violation, $\phi_s$ is the $B^0_s$--$\bar B^0_s$ 
mixing phase, and $\Delta\phi_{VV,k}$ is a non-perturbative hadronic phase shift. 
The expressions given in Ref.~\cite{LHCbKstarKstar} for the $B_s\to K^{*0}\bar K^{*0}$ decay 
take the leading order effect of $y_s$ into account,   
and assume $\phi_s=0$ and negligible hadronic corrections.

The generalization of Eq.~(\ref{BRratioT}) is given by 
\begin{equation}
	{\rm BR}_{\rm theo}^{VV}
	= \mbox{BR}_{\rm exp}^{VV}\sum_{k=0,\parallel,\perp} 
	\left[2 - \left(1-y_s^2\right) \frac{\tau_k^{VV}}{\tau_{B_s}}\right]
	f_{VV,k}^{\rm exp},	\label{BRratioT-VV}
\end{equation}
and does not require knowledge of the ${\cal A}_{\Delta\Gamma}^{VV,k}$ observables.

\section{Experimental Aspects}\label{sec:exp}
Additional subtleties arise in the experimental determination of $B_s$ branching ratios and effective 
lifetimes, in particular at a hadron collider environment
where many final-state particles are produced in the fragmentation.

Separating $B_s$ signal decays from the background typically involves selection criteria 
that use the flight distance of the $B_s$ meson or the impact parameter of its decay products,
leading to a decay-time dependent efficiency.
By rejecting short-living $B_s$ meson candidates, the relative
amounts of $B_{s,{\rm L}}$ and $B_{s,{\rm H}}$ mesons in the remaining data sample 
are altered, resulting in a biased result for the branching ratio determination.
The extrapolation of the event yield to full acceptance is usually obtained from simulation, 
but this requires {\it a priori} assumptions of the values for $y_s$ 
and ${\cal A}^f_{\Delta\Gamma}$. For example, the dependence of the branching fraction
correction on the value ${\cal A}^f_{\Delta\Gamma}$ can be several percent if only decay times greater than $0.5$~ps are considered.
This systematic uncertainty is avoided by tuning the simulation using the measured 
value of the effective lifetime.

Furthermore, the presence of remaining background events with a different observed decay time 
distribution as the signal, 
implies that it is experimentally unpractical to determine
the time expectation value $\tau_f$ of the untagged rate as given in  Eq.~\eqref{effLifetime}.
Instead, the effective lifetime is commonly extracted by fitting a single exponential 
function to the untagged rate \cite{Jpsif0,tauKK,HM}, which in general is described by two 
exponentials (see Eq.~\eqref{untagged}). 
In Appendix~\ref{sec:analytic} we demonstrate that such a fitting procedure leads to 
an unbiased determination of the effective lifetime in the case of a log likelihood fit and
to a small bias for a $\chi^2$ minimization procedure.

\section{Conclusions}
The established width difference of the $B_s$ mesons complicates the extraction of
branching ratio information from the experimental data, leading to biases at the $10\%$
level that depend on the specific final state. On the one hand, these effects can be
included through theoretical considerations and phenomenological analyses. On the other
hand, it is also possible to take them into account through the measurement of the
effective $B_s\to f$ decay lifetimes, which is the preferred avenue. So far, these effects
have not, or only partially, been included and we advocate to use the converted 
branching ratios for comparisons with theoretical calculations in particle listings.

\appendix
\section{Effective Lifetime Fits}\label{sec:analytic}
An {\it effective lifetime} for a $B_s$ decay channel is obtained in practice by
fitting a single exponential function to its untagged rate.  As an untagged rate is in general described by two exponentials, corresponding to two mass-eigenstates with different lifetimes, the single exponential fit is an approximation.

In order to find analytic expressions for the fitted effective lifetime
$\tau_{\rm eff}$, we let the untagged rate be the {\it true} Probability
Distribution Function (PDF), and the single exponent function the {\it fitted}
PDF, such that
\begin{align}
        f_{\rm true}(t) &\equiv \frac{A(t)\, 
        \langle\Gamma(t)\rangle}{\int^\infty_0 A(t)\,\label{PDFtrue} \langle\Gamma(t)\rangle\, dt},\\
        f_{\rm fit}(t;\tau_{\rm eff}) &\equiv \frac{A(t)\, 
        e^{-t/\tau_{\rm eff}}}{\int^\infty_0 A(t)\, e^{-t/\tau_{\rm eff}}\, dt},\label{PDFfit} 
\end{align}
where $A(t)$ is an acceptance efficiency function. 
The likelihood or $\chi^2$ function for the fit in question is then built using the above PDFs, 
and maximised or minimised, respectively, in the limit of infinitesimally spaced bins.
Specifically, for $n$ events we minimise the functions:
\begin{align}
        - \log L(\tau_{\rm eff}) =&\ - n \int_0^\infty dt\ f_{\rm true}(t)
                \log\left[ f_{\rm fit}(t;\tau_{\rm eff})\right], \\
        \chi^2(\tau_{\rm eff}) &= 
        n\int_0^\infty dt\,\frac{\left[ f_{\rm true}(t) - 
        f_{\rm fit}(t;\tau_{\rm eff})\right]^2 }{f_{\rm fit}(t;\tau_{\rm eff})},
\end{align}
for a maximum likelihood and a least squares fit, respectively.
In a modified least squares fit, where data is used to estimate the error, the denominator in the $\chi^2$ integrand should be replaced by $f_{\rm true}(t)$.
For the maximum likelihood fit, taking the infinitesimal bin limit is equivalent to an unbinned fit.

The effective lifetime $\tau_{\rm eff}$ resulting from these fits is then given implicitly by the formula:
\begin{equation}
        \frac{\int_0^\infty t\, e^{-t/\tau_{\rm eff}} \,A(t)\, dt}
        {\int_0^\infty  e^{-t/\tau_{\rm eff}}\,A(t)\, dt} = 
        \frac{\int_0^\infty t\, g(t;\tau_{\rm eff}) A(t)\, dt}
        {\int_0^\infty g(t;\tau_{\rm eff}) A(t)\, dt},
        \label{analyticGen}
\end{equation}
where
\begin{equation*}
        g(t;\tau_{\rm eff}) \equiv \left\{
        \begin{array}{lcl}
                \langle \Gamma(t)\rangle & : & {\rm maximum\ likelihood} \\     
                \langle \Gamma(t)\rangle^2\, e^{\,t/\tau_{\rm eff}} & : & {\rm least\ squares} \\
				\langle \Gamma(t)\rangle^{-1}\,	e^{-2\,t/\tau_{\rm eff}} & : & {\rm modified\ least\ squares.}  
        \end{array}
        \right.
\end{equation*}

The effective lifetime definition given in \eqref{effLifetime} is reproduced for 
the untagged rate given in \eqref{untagged2} 
if we assume a trivial acceptance function, $A(t)=1$, and apply a maximum likelihood fit.
For non-zero values of $y_s$, the least squares fits give different analytic 
expressions for the effective lifetime.
Fortunately, for the current experimental range of $y_s$, the differences are of the order 0.1\%.

\vspace*{-0.7truecm}
\section*{Acknowledgements} 
\vspace*{-0.4truecm}
We would like to thank the LHCb collaboration for discussions. 
This work is supported by the Netherlands Organisation for Scientific Research (NWO)
and the Foundation for Fundamental Research on Matter (FOM).


%
%
%
\end{document}